\documentclass[review]{elsarticle}
\usepackage{graphicx}
\usepackage{dcolumn}
\usepackage{bm}
\usepackage{empheq,etoolbox}
\usepackage{mathtools}
\usepackage{grffile}
\usepackage{subcaption}
\usepackage{breqn}
\usepackage{tikz}
\usepackage{float}
\usepackage[labelsep=period]{caption}
\usetikzlibrary{calc}
\usetikzlibrary{arrows}
\usepackage{SIunits}
\usepackage{pgfplots}
\usetikzlibrary{pgfplots.groupplots}
\usepackage{amsmath,latexsym}
\usepackage{xspace}

\usepackage{lineno,hyperref}
\modulolinenumbers[5]

\journal{Journal of Applied Surface Science}

\begin{document}

\begin{frontmatter}

\title{\textbf{Theory for heating of metals assisted by Surface Plasmon Polaritons}}
\author[mymainaddress]{O. Benhayoun\corref{mycorrespondingauthor}}
\ead{obenhayoun@uni-kassel.de}
\author[mysecondaryaddress]{P. N. Terekhin}
\author[mytertieryaddress]{D. S. Ivanov}
\author[mysecondaryaddress]{B. Rethfeld}
\author[mymainaddress]{M. E. Garcia}
\address[mymainaddress]{Institute of Physics and Center for Interdisciplinary Nanostructure Science and Technology (CINSaT)$,$ University of Kassel$,$ 34132 Kassel$,$ Germany}
\address[mysecondaryaddress]{Department of Physics and Research Center OPTIMAS$,$ Technische Universität Kaiserslautern$,$ 67663 Kaiserslautern$,$ Germany}
\address[mytertieryaddress]{Lebedev Physical Institute$,$ 119991 Moscow$,$ Russia}

\cortext[mycorrespondingauthor]{Corresponding author at: Institute of Physics and Center for Interdisciplinary Nanostructure Science and Technology (CINSaT)$,$ University of Kassel$,$ 34132 Kassel$,$ Germany}

\begin{abstract}
We propose a model accounting for plasmons within a two temperature description, to investigate the role of surface plasmon polaritons (SPP) in the energy redistribution between laser excited electrons and the lattice, leading to surface restructuring of the material. This energy transfer can lead to the creation of laser induced surface structures in metals illuminated by an ultrashort laser pulse. The Two Temperature Model + Plasmon (\textrm{TTM+P}) equations are constructed by applying perturbation theory on the energy-, momentum- and density conservation equations of free electrons in metals.  We consider three subsystems: the lattice, the thermalized electrons and the SPP, subject to an external laser field irradiation. The interference between the laser and SPP fields leads to spatially modulated energy absorption by the electronic system, and, through electron-phonon collisions, periodically shapes the lattice temperature. A numerical analysis is performed on a 1D model for gold. We show the emergence of a periodic modulation of the lattice temperature, which may contribute to laser induced periodic surface structures  (\textrm{LIPSS}). 
\end{abstract}

\begin{keyword}
Ultrashort laser pulses, surface Plasmon Polaritons, Hydrodynamic model, TTM, LIPSS.
\end{keyword}

\end{frontmatter}

\section{Introduction}
Surface Plasmon Polaritons (SPP) are collective electronic excitations coupled with light. They motivate considerable interest in the scientific community for both their theoretical and technological aspects, as they provide a wide spectrum of applications, ranging from biosensing \cite{chien2004sensitivity, pitarke2006theory}, surface wetting \cite{herminghaus1997hydrogen}, sub-wavelength optics \cite{barnes2006surface, luo2012surface} and laser induced periodic surface structures (LIPSS) \cite{bonse2009role, gurevich2014laser, bonsemaxwell}. SPP can only be excited when their momentum matches that of the incoming light. Thus, for an ideal planar metal surface, this cannot be achieved, and one has to provide the requisite momentum to the laser via attenuated total reflection \cite{otto1968excitation, kretschmann1968radiative}, or via surface roughness or gratings \cite{pitarke2006theory,   bonsemaxwell, terekhin2019influence, maier2007plasmonics}. Consequently, the electromagnetic fields associated with SPP are confined to the surface, and decay outside and inside the material, hence propagating along the metal-dielectric interface until their energy is dissipated as heat. 
\\ \indent In order to understand the influence of the electromagnetic fields of SPP on the lattice temperature distribution, we use the phenomenological Two Temperature Model (\textrm{TTM}) \cite{anisimov1974electron, singh2010two, rethfeld2017modelling}. This model is widely applied to study the energy dissipation inside the material after an ultrafast laser pulse excitation. The external laser field drives the electronic and lattice systems out of thermal equilibrium, subsequently the subsystems exchange energy via  electron-phonon coupling. This is described with the following equations:
\begin{subequations} \label{TTM}
  \begin{empheq}{gather}
    C_e \partial_t T_e + \nabla\cdot \textbf{Q}_e = -G_{\textrm{e-ph}}(T_e - T_\ell) + S_e, \\
    C_\ell \partial_t T_\ell + \nabla\cdot \textbf{Q}_\ell= G_{\textrm{e-ph}}(T_e - T_\ell)\textrm{.} \label{ttm lattice}
  \end{empheq}
\end{subequations}
where $T_e$ and $T_\ell$ are the electronic and lattice temperatures, $\textbf{Q}_e$ and $\textbf{Q}_\ell$ are the heat fluxes, $G_{\textrm{e-ph}}$ is the electron-phonon coupling parameter and $S_e$ is the source term, which describes laser energy absorbed by the electronic system. \\
The inclusion of SPP fields can be done by modifying the source term $S_e$ using Maxwell's equations  \cite{terekhin2019influence, gurevich2017role, shih2020effect, levy2016relaxation, levy2017laser, shugaev2017mechanism, gurevich2020three}. This is a quite useful approach in the context of studying the contribution of SPP fields in the LIPSS formation. In this case, the electronic temperature distribution is spatially modulated by the interference of SPP with the laser source. Then, as a result of the electron-phonon relaxation, this spatial modulation is transferred to the lattice temperature, hence inducing surface topography modifications via ablation and material transport, creating ripples \cite{bonsemaxwell, terekhin2019influence, bonse2016laser, gurevich2016mechanisms}. However, the derivation of the new source term often relies on arbitrary parameters to complete the description of the SPP fields, and sometimes even impose arbitrary spatial distributions \cite{gurevich2017role, shih2020effect, levy2016relaxation, levy2017laser, shugaev2017mechanism, gurevich2020three}. Recently, it was shown that the number of free arbitrary parameters can be reduced down to only two  \cite{terekhin2019influence}. \\
\indent In this paper, we propose a new approach for deriving the source term in Eq. (\ref{TTM}), by describing the SPP as an independent degree of freedom, using the hydrodynamic model. For this, we write the energy-, momentum- and mass- conservation equations for the electronic system, and linearize them via perturbation theory. We end up with a coupled system of equations that describes the electronic collective excitations, as well as the energy exchange between SPP, thermalized electrons and the lattice. Note that the topographical and the lattice re-organization effects are not included and are beyond the scope of this letter. We find from our 1D simulations of a gold sample, that upon irradiation with an ultrafast laser pulse, we obtain an oscillatory pattern in the lattice temperature, whose periodicity is then compared with predictions from an analytical dielectric function of Au.

\section{The \textrm{TTM+P} model}\label{theory}
\subsection{Conservation laws}

To develop a model describing collective electronic excitation without making arbitrary constraints on the shape of the SPP fields, we start by obtaining the conservation equations governing our electronic system. We then follow a similar approach as in the work of Chen \textit{et al.} \cite{chen2006semiclassical} to derive the TTM+P equations.
\\\indent We describe a nonequilibrium system of free electrons irradiated by an external laser field using the moments of the Boltzmann equation for mass, momentum and energy, which reads:
\begin{subequations} \label{conservation eq}
  \begin{empheq}[]{gather}
    \partial_t n + \nabla \cdot (n \textbf{v}) = 0, \\
    m D_t \textbf{v} = e \textbf{F}_e - \frac{1}{n} \nabla p, \label{momentum conservation}\\
    D_t \xi + \frac{1}{n}(p\nabla\cdot\textbf{v} + \nabla\cdot\textbf{Q}) = (\partial_t\xi)_c\textrm{.}
  \end{empheq}
\end{subequations}

\noindent with $n$ the electronic particle density, $\textbf{v}$ the mean velocity vector, $\xi$ the internal energy, $p$ the electronic pressure and the energy flux vector $\textbf{Q}$, $\textbf{F}_e = e (\textbf{E} + \textbf{v} \times \textbf{B} )\ $ the Lorentz force resulting from the electromagnetic field \textbf{E} and \textbf{B}, $(\partial_t \xi)_c$ the collision term, $\partial_t = \frac{\partial}{\partial t}$ and $D_t = \partial_t + \textbf{v}\cdot\nabla$.
\\\indent Note that the above equations are coupled via $n, \textbf{v}$ and $\textbf{F}_e$ to the microscopic Maxwell's equations:

\begin{subequations} \label{maxwell}
  \begin{empheq}[]{gather}
    \nabla \times \textbf{B} = \mu_0\textbf{J} + \frac{1}{c^2}\partial_t \textbf{E},\\
    \nabla \times \textbf{E} = - \partial_t \textbf{B},\\
    \nabla \cdot \textbf{B} = 0,\\
    \nabla \cdot \textbf{E} = \frac{e}{\epsilon_0} (n-l)\textrm{.}
  \end{empheq}
\end{subequations}

with
$\epsilon_o$ and $\mu_0$ are the permittivity and permeability of free space respectively, $l=\sum_i Z_i \delta(\textbf{r}-\textbf{r}_i)$ is the distribution function of the positive ion background, $Z_i$ being the effective charge of the ions and $\textbf{r}_i$ is the position of the $i$-th ion, and the electronic current $\textbf{J}=en\textbf{v}$.
\\These two sets of equations are the basis for the derivation of our model.

\subsection{Perturbation equations}

\indent From the system of Eqs. (\ref{conservation eq}), we are able to describe collective electronic excitations. We thus use the perturbation theory on the Hamiltonian of the electronic system to split it into two parts: a “non-perturbed” zero-th order system, describing a uniform and static electronic background (i.e. $n_0 (\textbf{r}, t) = n_0$ and $\textbf{v}_0 (\textbf{r}, t) = \textbf{0}$, respectively), and a “perturbed” system to the first order consisting of plasmons, which provides a linear correction to the zero-th order equations, and hence describes the deviations of the electronic density from uniformity. 
\\ We now expand our variables as follow  \cite{pitarke2006theory, ritchie1969photon, wakano1961foundations, ritchie1962optical}:
\begin{subequations} \label{perturbation}
  \begin{empheq}[]{gather} 
    n(\textbf{r},t) = n_0 + n_1(\textbf{r},t), \\
    \textbf{v}(\textbf{r},t) =  \textbf{v}_1(\textbf{r},t),\\
    p(\textbf{r},t) = p_0(\textbf{r},t) +  p_1(\textbf{r},t),\\
    \textbf{E}(\textbf{r},t) = \textbf{E}_0(\textbf{r},t) +  \textbf{E}_1(\textbf{r},t),\\
    \textbf{B}(\textbf{r},t) =  \textbf{B}_1(\textbf{r},t),\\
    \xi(\textbf{r},t) = \xi_0(\textbf{r},t) + \xi_1(\textbf{r},t)\textrm{.}
\end{empheq}
\end{subequations}

The linear deviations are only small corrections to the unperturbed system, and satisfy the inequality:
\begin{equation}\label{perturbation_condition}
    n_1(\textbf{r},t) << n_0\textrm{.}
\end{equation}
which assumes that the number of excited electrons is small compared to the total number
of electrons in the system. This perturbative approach gives us approximate solutions to an otherwise, difficult problem to solve.
We now insert (\ref{perturbation}) into equations  (\ref{conservation eq}) and (\ref{maxwell}). After separating the zero-th and first order variables, we find:
\begin{subequations} \label{conservation eq01}
  \begin{empheq}[]{gather} 
    D_t \xi_0 + \frac{1}{n_0}\nabla\cdot\textbf{Q}_e = (\partial_t\xi_0)_c,  \label{conservation eq01_1} \\
    \partial_t n_1 + n_0\nabla \cdot \textbf{v}_1 = 0, \\
    \nabla \cdot \textbf{E}_1 = \frac{e}{\epsilon_0} n_1, \\
    \partial_t\textbf{v}_1 +\frac{\beta^2}{n_0}\nabla n_1 = \frac{e}{m} \textbf{E}_1  \label{conservation momentum},\\
       n_0 \partial_t \xi_1 + n_1 \partial_t \xi_0 + n_0 \textbf{v}_1 \cdot \nabla \xi_0 + p_0 \nabla \cdot \textbf{v}_1 = (\partial_t\xi_1)_c, \label{conservation eq01_5} \\
    \nabla \times \textbf{B}_1 = \mu_0\textbf{J}_1 + \frac{1}{c^2}\partial_t \textbf{E}_1\textrm{.}
\end{empheq}
\end{subequations}

\noindent where we have used the condition for the conservation of the generalized vortex $\nabla \times \textbf{v}_j + \frac{e}{m}\textbf{B}_j = \textbf{0}$ for $j \in \{0, 1\}$, $n_0$ is the uniform electron density and $\textbf{v}_0 = \textbf{0}$. We also replaced in Eq. (\ref{perturbation}.c), the first order pressure term by the degeneracy Fermi pressure and is written as $p_1= mn_1\beta^2$ with $\beta=\frac{(3\pi^2 n_0)^{1/3} \hbar}{m}$ the perturbation velocity \cite{pitarke2006theory, ritchie1969photon}. The model can be improved by making $\beta$ a function of position as well as of plasmon frequency \cite{kleinman1973improved}, however this is beyond the scope of our study. Finally, the electric current $\textbf{J}_1 =en_0\textbf{v}_1$. We also wish to indicate here, that the fields $\textbf{E}_1$ and $\textbf{B}_1$ are already coupled with light, and can be decomposed as $\textbf{E}_1 = \textbf{E}_{\textrm{sp}} + \textbf{E}_{\textrm{las}}$ and $\textbf{B}_1 = \textbf{B}_{\textrm{sp}} + \textbf{B}_{\textrm{las}}$, with $\textbf{E}_{\textrm{sp}}$ and $\textbf{B}_{\textrm{sp}}$ the electromagnetic fields of surface plasmons in the absence of external irradiation. 
\\ We note that the condition (\ref{perturbation_condition}), can be re-written, using Eq. (\ref{conservation eq01}.d) as:
\begin{equation}\label{inequality}
    e\phi_1 + m \partial_t s_1 << m\beta^2\textrm{,}
\end{equation}
where $\phi_1$ and $s_1$ are the electric and velocity potentials respectively, which are related to $\textbf{E}_1$ and $\textbf{v}_1$ with: $\textbf{E}_1 = -\nabla \phi_1 - \partial_t \textbf{A}_1$ and $\textbf{v}_1 = - \nabla s_1 + e\textbf{A}_1/mc$, $\textbf{A}_1$ being the vector potential in the Coulomb gauge. The left hand side of the Eq. (\ref{inequality}) represents the sum of the potential electric energy and an effective potential, and the right hand side can be described as the kinetic energy of the perturbation.\\
\indent In order to be able to describe SPP, we impose on the system of Eqs. (\ref{conservation eq01}) the following boundary conditions \cite{pitarke2006theory, ritchie1969photon, wakano1961foundations}:\\
\indent 1- Continuity of the normal and tangential components of the electromagnetic field across the interface. \\
\indent 2- The normal component of the electronic velocity $v_{1z}$ vanishes at $z=0$.

\subsection{Energy balance} \label{energy balance}
From Eqs. (\ref{conservation eq01_5}, \ref{conservation eq01_1}, \ref{ttm lattice}), we finally end up with \cite{chen2006semiclassical}:
\begin{subequations} \label{TTM+P}
  \begin{empheq}[]{gather}
 \begin{split}
        n_0 \partial_t \xi_1 + n_1 C_e \partial_t T_e + n_0 C_e (\textbf{v}_1 \cdot \nabla)T_e \\ +\frac{2}{3}C_e T_e \nabla\cdot\textbf{v}_1 = (\xi_1)_c,
\end{split}\\ 
    C_e \partial_t T_e + \nabla\cdot \textbf{Q}_e = (T_e)_c + S_e, \\
    C_\ell \partial_t T_\ell + \nabla\cdot \textbf{Q}_\ell=(T_\ell)_c\textrm{.} 
  \end{empheq}
\end{subequations}

\noindent Eqs. (\ref{TTM+P}) describe the energy balance of the SPP, the thermalized electrons and the lattice subsystems respectively, where $\textbf{Q}_i = -K_i \nabla T_i$ is the heat flux vector for ($i = e$, $\ell$) and $S_e = -\nabla \cdot \textbf{E}_{\textrm{tot}} \times \textbf{B}_{\textrm{tot}}$ is the source term describing the interference of the SPP fields with the laser fields, with $\textbf{E}_{\textrm{tot}} = (\textbf{E}_1+\textbf{E}_{\textrm{las}})$ and $\textbf{B}_{\textrm{tot}} = (\textbf{B}_1+\textbf{B}_{\textrm{las}})$. The Bhatnagar–Gross–Krook (\textrm{BGK}) model \cite{bhatnagar1954model} has been used for the collision terms $(\xi_1)_c$, $(T_e)_c$ and $(T_\ell)_c$. We thus have:
\begin{subequations} \label{collision Terms}
  \begin{empheq}[]{gather}
    (\xi_1)_c = - G_{\textrm{e-pl}}(\frac{\xi_1}{C_e} - T_e), \\ 
    (T_e)_c = -G_{\textrm{e-ph}}(T_e - T_\ell)+ G_{\textrm{e-pl}}(\frac{\xi_1}{C_e} - T_e), \\
    (T_\ell)_c =  G_{\textrm{e-ph}}(T_e - T_\ell)\textrm{.} 
  \end{empheq}
\end{subequations}
with $G_{\textrm{e-ph}}$ and $G_{\textrm{e-pl}}$ are the electron-phonon and electron-plasmon coupling parameters, respectively.

\section{Numerical simulations}\label{numsym}
\begin{figure}[H]
\includegraphics[scale=.5]{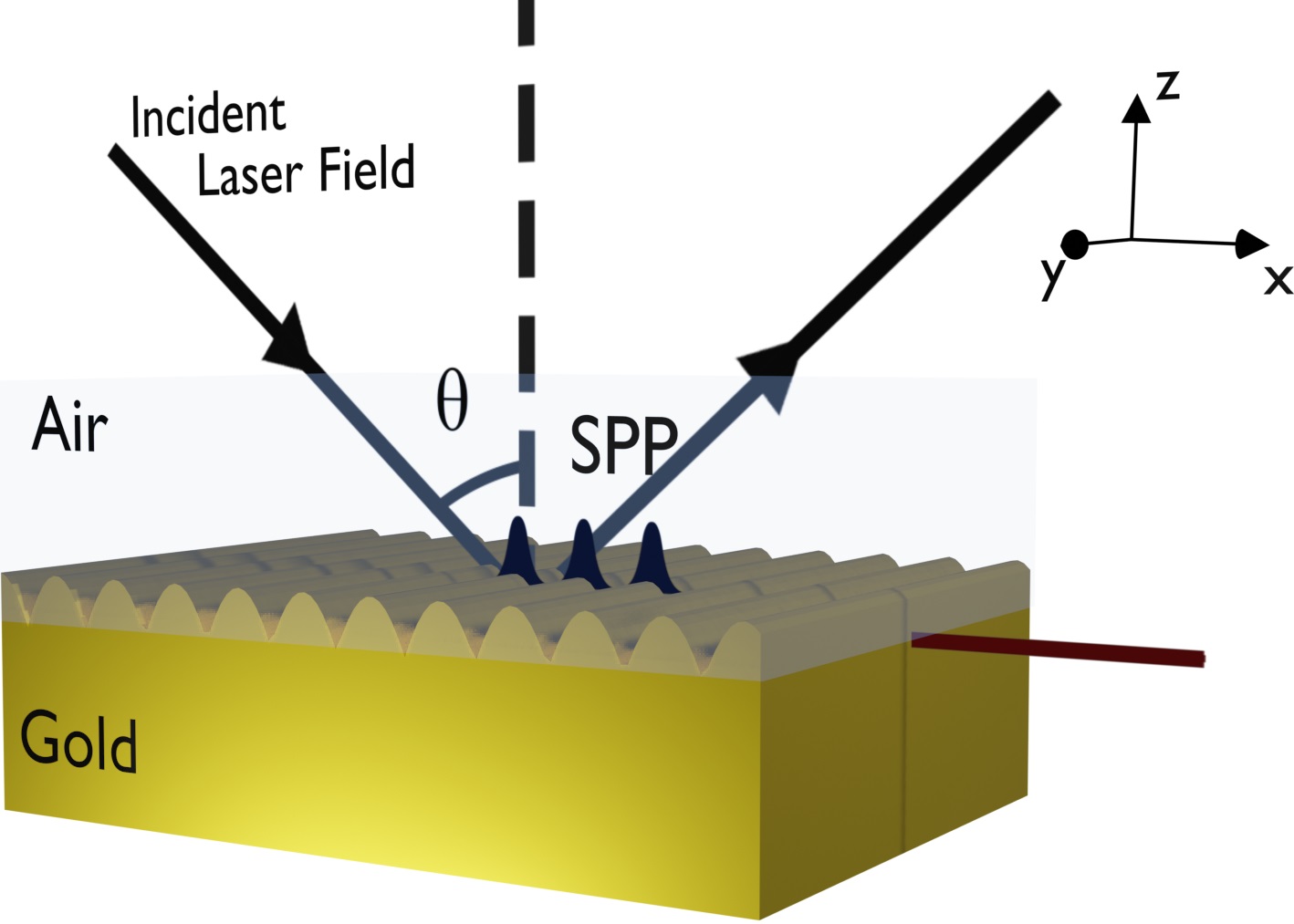}
\caption{Geometry of the gold sample used in our simulations. We reduce the system to a 1D problem, ~ by supposing homogeneity in the $y$-direction and considering only the surface of the material (at $z=0$). At $t=0$, we excite SPP using an obliquely incident laser field with an angle $\theta$ and the surface of the material has no predefined structures. During the simulations, SPP start to emerge and contribute to the appearance of periodic surface structures at the gold-air interface.}
\label{geometry11}
\end{figure}

In this section, we perform a numerical analysis of the system of equations (\ref{conservation eq01}-\ref{TTM+P}).  Fig.(\ref{geometry11}) shows the geometry of the modeled sample. We have reduced the 3D problem into 1D by supposing homogeneity in the y-direction and focusing on the surface modes of the plasmons at $z=0$. We note that in this case, the penetration and damping of \textrm{SPP} inside and outside of the material is beyond the scope of this letter. 

\subsection{Laser's electromagnetic field}

The laser's electromagnetic field at $z=0$ is described by the real part of:
\begin{subequations}\label{laser field}
  \begin{empheq}{gather}
    E^x_{\textrm{las}}(x,t) = E_0^{\textrm{abs}}\cos(\theta_a)\xi_{\textrm{las}}(t)e^{-i(k_0  \sin(\theta_i) x + \omega t)},\\
    E^z_{\textrm{las}}(x,t) =-E_0^{\textrm{abs}}\sin(\theta_a)\xi_{\textrm{las}}(t)e^{-i(k_0  \sin(\theta_i) x + \omega t)},\\
    B^y_{\textrm{las}}(x,t) = B_0^{\textrm{abs}}\xi_{\textrm{las}}(t)e^{-i(k_0 \sin(\theta_i) x + \omega t)}\textrm{.}
  \end{empheq}
\end{subequations}
with
\begin{equation}
\begin{split}
    \xi_{\textrm{las}}(t) = \sqrt{\frac{1}{\tau}\frac{\sqrt{\sigma}}{\sqrt{\pi}}}\exp\left({-\frac{\sigma t^2}{2\tau^2}}\right)\textrm{,}
    \end{split}
\end{equation}
where $\omega$ and $k_0 = \omega /c $ are the laser frequency and light wave vector respectively, the parameter $\sigma = 4 \ln{(2)}$. The laser pulse duration is $\tau$. $E_0^{\textrm{abs}} = \frac{2 \eta \cos(\theta_a)}{ \eta \cos(\theta_a) + \eta_0 \cos(\theta_i)}E_0^{\textrm{inc}}$ and $B_0^{\textrm{abs}}= -\Tilde{n}E_0^{\textrm{abs}}/c$ are the amplitude of electric and magnetic fields inside the material respectively, $\eta = \sqrt{\frac{\mu}{\epsilon}}$ and $\eta_0 = \sqrt{\frac{\mu_0}{\epsilon_0}}$ are the material and vacuum impedance respectively, $\theta_a = \arcsin(\sin(\theta_i)/\Tilde{n})$ is the angle of transmittance related to the angle of incidence $\theta_i$, $\Tilde{n}$ is the material complex refractive index, $E_0^{\textrm{inc}}$ is the incident amplitude of the laser electromagnetic field. 
The fields in Eq. (\ref{laser field}) are used to derive the SPP fields, and to calculate the source term $S_e$ in Eq. (\ref{TTM+P}.b). We note that in the case of non-flat surfaces (periodic grating or roughness, for instance), one should use analytical expressions for the laser field that account for its scattering on the specific surface topography. Hence, we would suggest changing the absorbed and reflected laser fields \cite{oladyshkin2019optical, maradudin1975scattering}. We note that in the case of multiple pulses, the model does not provide a description of the lattice surface modification and thus fails to take the interpulse feedback mechanism into consideration. This however, goes beyond the scope of our study.

\subsection{Algorithm of the numerical simulation}
In 1D, our system of equations reads as follow:
\begin{subequations} \label{conservation eq16}
  \begin{empheq}[]{gather}
     \Box_c\textbf{B}_1 = \textbf{0}, \\
    \Box_{\beta} n_1= 0, \\
    \Box_c \textbf{E}_1 = \frac{e}{\epsilon_0}(\beta^2 - c^2)\partial_x n_1 \textbf{e}_x,\\
    \Box_{\beta}\textbf{v}_1= -\frac{e}{m}((\beta^2 - c^2) \nabla \times \textbf{B}_1 \cdot \textbf{e}_z)\textrm{.} %
  \end{empheq}
\end{subequations}
with the operators $\Box_c = \partial_t^2 - c^2\partial_x^2 + \omega_p^2  + \frac{\partial_t}{\tau_{\textrm{ep}}}$ and $\Box_{\beta} =   \partial_t^2 - \beta^2\partial_x^2 + \omega_p^2 + \frac{\partial_t}{\tau_{\textrm{ep}}}$, $\textbf{e}_x$ and $\textbf{e}_z$ are the unit vectors, and $\tau_{\textrm{ep}}$ is the electron-plasmon relaxation time.\\
We start by solving the system of Eqs. (\ref{conservation eq16}) for the SPP fields at each time step, and the values of these calculated variables are used to solve Eqs. (\ref{TTM+P}).
Along with the diffusion equations, the wave equations were solved using a central finite difference scheme. 
\\A uniform mesh of grid spacing of 4 \textrm{nm} size was employed for a total length of $L=4~\mu\textrm{m}$, along with a time increment of $5 \cdot 10^{-18}$ \textrm{s}, for a duration of 50 \textrm{ps}.
 As for the boundaries, reflective boundary conditions are used for the wave equations, so that the variables in Eqs. (\ref{conservation eq16}) satisfy $f(x,t) = 0$ at $x = 0$ and $x = L$, and convective boundaries were adopted for the diffusion equation, i.e. $\partial_xT_{e,l} = 0$. 

\subsection{Material parameters}\label{material parammeters}
The material parameters that were used are as follows: $n_0 =  5.9 \cdot 10^{28}  \textrm{m}^{-3}$ \cite{fourment2014experimental}, and we write $G_{\textrm{e-pl}} = C_e/\tau_{\textrm{ep}}$, with $\tau_{\textrm{ep}}=\frac{\hbar\omega_p}{(\hbar^2k_0^2/m)}(\frac{k^2}{k_0^2})\omega_p$ the SPP lifetime due to electron-plasmon collisions \cite{pines1956collective,nozieres1959electron}, $k_0$ being the momentum of an electron at the top of the Fermi distribution and $k$ the SPP wave vector. Moreover, $G_{\textrm{e-ph}}(T_e)$ and $C_e(T_e)$ were taken from the work by Lin \textit{et al.} \cite{lin2006thermal}. The electronic thermal conductivity $k_e$ was approximated as in \cite{anisimov1997theory}: $k_e = K\frac{(a_e^2+0.16)^{5/4}(a_e^2 + 0.44)a_e}{(a_e^2 + 0.092)^{1/2}(a_e^2+ba_\ell)}$, with $a_i = k_B T_i/ \epsilon_F$ ($i \in \{e,l\}$ and $\epsilon_F$ is the Fermi energy) and $K$ and $b$, are constants depending on the material, here taken as $K= 353$ $\textrm{W}\textrm{m}^{-1}\textrm{K}^{-1}$ and $b= 0,16$ for gold, and $k_\ell = 1$ $\textrm{W} \textrm{m}^{-1} \textrm{K}^{-1}$ \cite{wang2016first}. We note that these parameters do not take into account phase transitions or boiling processes of the lattice, since they do not influence the periodicity emerging in the temperature distribution, which is the main focus of our calculations. Also, the ballistic transport of the electrons can be omitted for the range of laser fluence used in our simulations \cite{petrov2015two}. 
\\ Finally, the initial conditions are $T_e(x, t=0)=T_\ell(x, t=0)=T_{in}=293$K and $\epsilon_1(x,t=0)=C_eT_{in}$, and for the SPP fields, the initial conditions were derived from the laser electromagnetic field at the interface as: \begin{subequations} \label{initial conditions}
  \begin{empheq}[]{gather}
    E_{1x}(x,t=0) = E^x_{\textrm{las}}(x,t=0), \\
    E_{1z} (x,t=0) = E^z_{\textrm{las}}(x,t=0), \\
    B_{1y} (x,t=0) = B^y_{\textrm{las}}(x,t=0), \\ 
    n_1(x,t=0) = \frac{\epsilon_0}{e} \partial_x E_{1x}|_{t=0}, \\
    v_{1x}(x,t=0) = -\frac{\epsilon_0}{e n_0} \partial_t E_{1x}|_{t=0},\\ 
    v_{1z}(x,t=0) = 0 \textrm{.}
  \end{empheq}
\end{subequations}

\section{Results and discussion}\label{results}
We now present the results of our simulations, performed for gold and a laser pulse of different wavelengths and angles of incidence.
The goal of these simulations is to model the influence of the electromagnetic field of SPP on the lattice temperature distribution. 
\begin{figure}[H]
  \centering
  \includegraphics[width=1\linewidth]{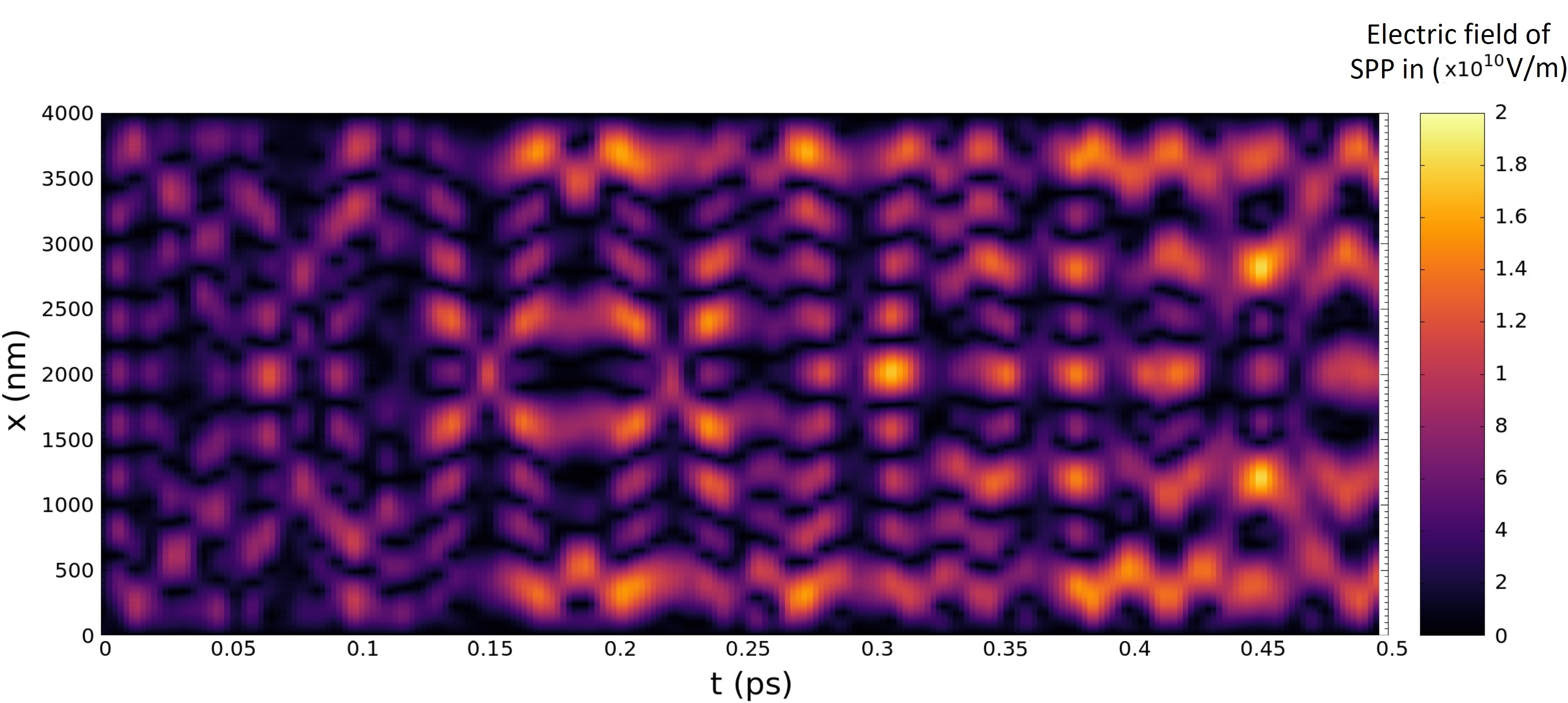}
   \caption{Space and time evolution of the electric field of the SPP.}  
  \label{Emfields}
\end{figure}

Fig. \ref{Emfields} displays the evolution in space and time of the electric field of the SPP. One can see at $t = 0$ the field's distribution due to the imposed initial conditions in Eqs. (\ref{initial conditions}), and how the field progresses in time. These initial conditions have a great influence on the SPP modes, propagation and their amplitudes, and have been chosen in such a way to remove arbitrary parameters in our model. This choice can be regarded as being in the ideal case where one has a perfect coupling between the laser and the surface plasmon fields. The intensification of the SPP eventually becomes clearer as the laser reaches its peak at around $t = 0\textrm{.}25$ ps. As mentioned before, this will then lead to an interference between the SPP fields and the laser fields, which will introduce a periodic distribution in the lattice temperature as a result of the electron-phonon relaxation \cite{bonse2009role,bonsemaxwell,colombier2012effects, colombier2020mixing,skolski2012laser, skolski2014modeling,zhang2015coherence,zhang2020laser, nakhoul2021self}. 
\begin{figure}[H]
  \centering
  \includegraphics[width=1\linewidth]{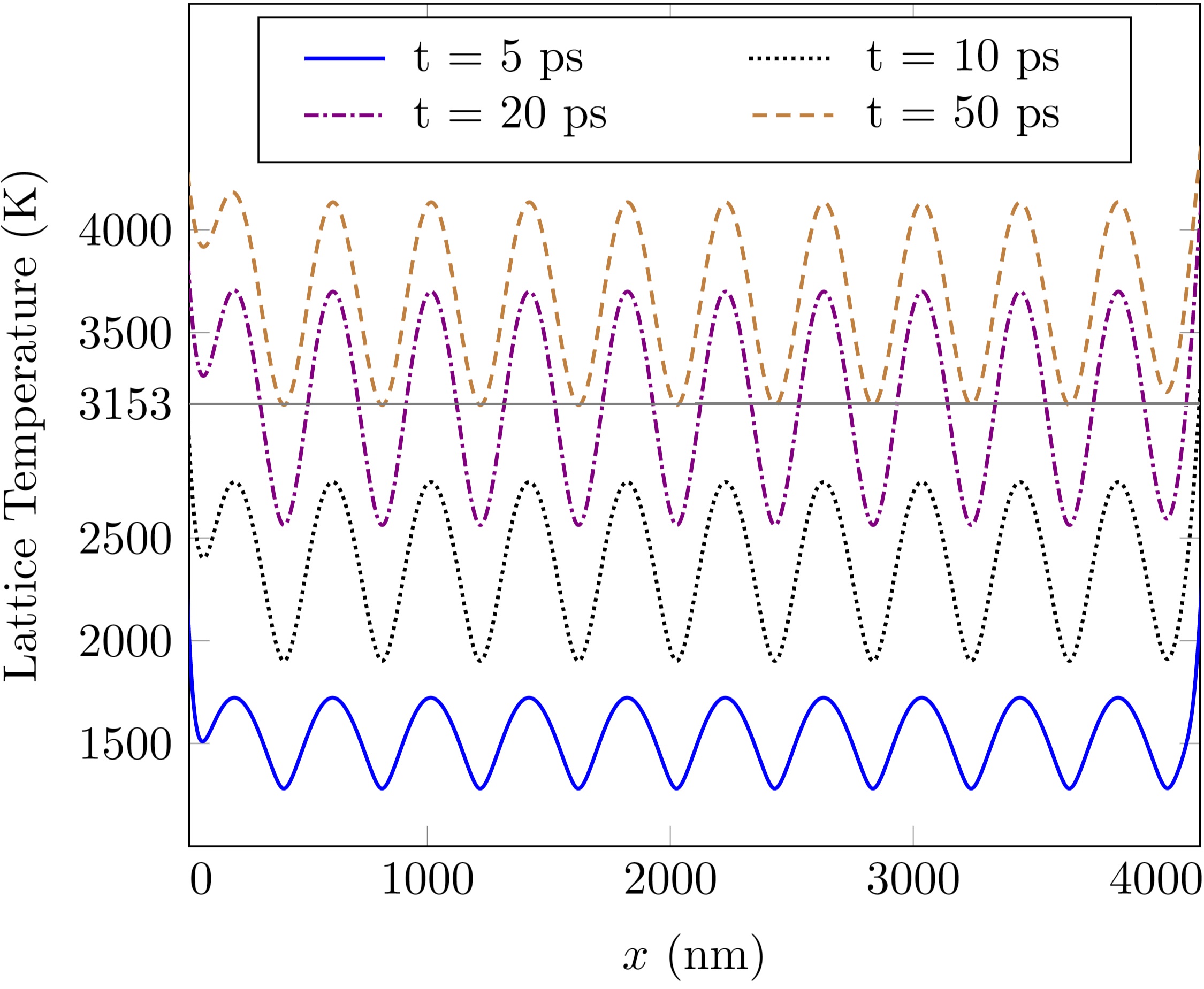}
   \caption{Time evolution of the lattice temperature distribution in a $4$ $\mu$m 1D gold sample irradiated by a laser pulse incident at $\theta = 60$\textdegree, with a 250 \textrm{mJ/cm}$^2$ incident fluence at 700 \textrm{nm} wavelength and a 100 \textrm{fs} duration . The gray line indicates the ablation temperature of gold.}  
  \label{timeEvo}
\end{figure}
\indent In Fig. \ref{timeEvo} we show the evolution of the lattice temperature of the sample which has been irradiated by a single pulse of 250 \textrm{mJ/cm}$^2$. We can see from this picture that around $t = 10~\textrm{ps}$, some peaks from the lattice temperature already reach the melting threshold of gold ($T_{\textrm{melting}} = 1337$ K). At $t = 50~\textrm{ps}$,  the lattice temperature increase becomes negligible, and its peaks exceed the ablation threshold, leading to a periodic material removal and transport, which after solidification, results in structural changes at the surface of the sample. In order to be able to see the complete picture, one has to incorporate matter reorganization theories into the model to accurately describe the surface topography evolution, since these would include hydrodynamic effects of the melted surface, diffusion and defects accumulation. However, for the purposes of this study, these effects are not taken into account, since they only play an important role after longer periods of time ($>1$ ns). 
\begin{figure}[H]
  \centering
  \includegraphics[width=.9\linewidth]{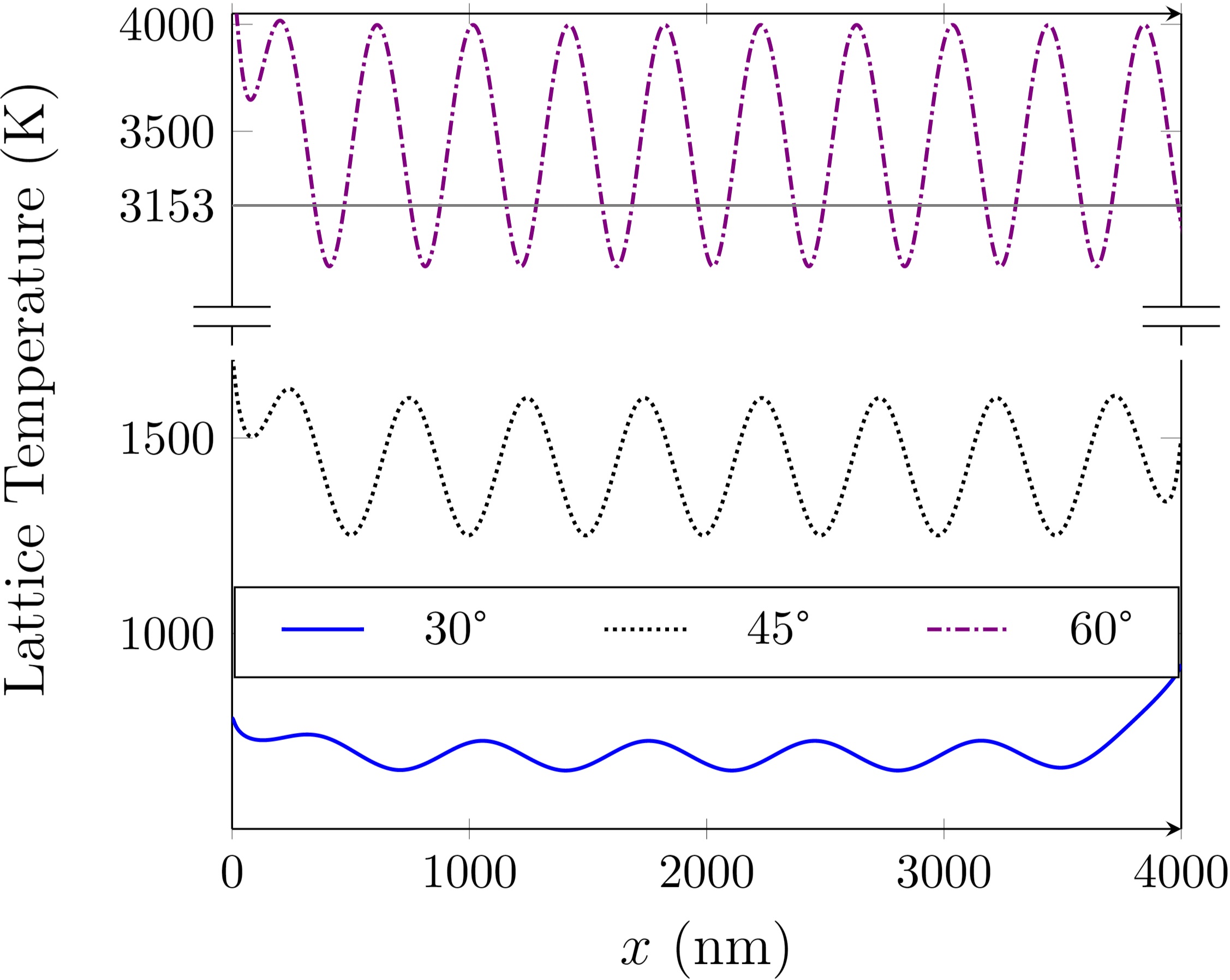}
   \caption{Lattice temperature distributions of the gold sample at $t = 30$ ps irradiated by laser pulses with different angles of incidence, with a 250 \textrm{mJ/cm}$^2$ incident fluence at a 700 \textrm{nm} wavelength and a 100 \textrm{fs} duration. The dashed gray line indicates the ablation temperature of gold.}
  \label{angles}
\end{figure}
\indent Fig. \ref{angles} demonstrates that, depending on the angle of incidence, we give to the SPP fields different amounts of energy. This can be seen from the temperatures reached by the lattice, and also by the amplitude of its oscillations. The closer we get to a grazing incidence, the greater the normal component of the laser field becomes, and hence the stronger we excite the SPP fields. We note, that with this particular implementation of initial conditions, one cannot excite SPP with a normal incidence, thus one would need to describe the initial state (density and velocity) of the electronic system independently of the laser source, to then be able to couple the surface plasmon fields and laser. This however, is beyond the scope of this study. We note that the increase of the lattice temperatures at the boundaries in Fig. \ref{timeEvo} and \ref{angles} is merely an artefact induced by the reflective boundary conditions used for solving the wave equations. 
Unfortunately, we are not aware of other papers that use the TTM+Maxwell approach for a 1D gold sample to have a reasonable comparison with. However, a Maxwell + TTM + Hydrodynamics model was reported for stainless steel in \cite{rudenko2019amplification}, and combined TTM and Maxwell simulations were performed in other studies on dielectrics and semiconductors (see \cite{bogatyrev2011non, hallo2007model, mezel2008formation}).

In order to investigate the SPP periodicity predicted by the hydrodynamic model, one can write its dielectric function using the Fourier transform of Eq. (\ref{conservation eq01}.f) and the condition $\nabla \times \textbf{v}_1 + \frac{e}{m}\textbf{B}_1 = \textbf{0}$, as:
\begin{equation}
    \epsilon_{\textrm{Hydro}} = 1 - \frac{\omega_p^2}{\omega^2 + i\omega/\tau_{\textrm{ep}} - \beta^2k^2},
\end{equation}
which when $\beta \longrightarrow 0$, is simply the dielectric function from the local Drude model. We recall that the dispersion relation of the SPP obtained using the interface conditions of the Maxwell's equation can be written as \cite{pitarke2006theory}:
\begin{equation}
    k_{\textrm{SPP}} = k_0\sqrt{\frac{\epsilon_{air}\epsilon_{Au}}{\epsilon_{air}+\epsilon_{Au}}},
\end{equation}
where $\epsilon_{air}$ and $\epsilon_{Au}$ are the dielectric constants for air and gold, respectively. From this relation, one can calculate the predicted wavelength of the SPP at an oblique incidence propagating in the positive direction of $x$, using the relation \cite{bonsemaxwell}: \begin{equation}\label{lambdaSPPs}
\lambda_{\textrm{SPP}} = \frac{2\pi}{Re(k_{\textrm{SPP}}) + k_0 sin(\theta_a)}\textrm{.}
\end{equation}
\begin{figure}[H]
 \centering
    \includegraphics[width=1\linewidth]{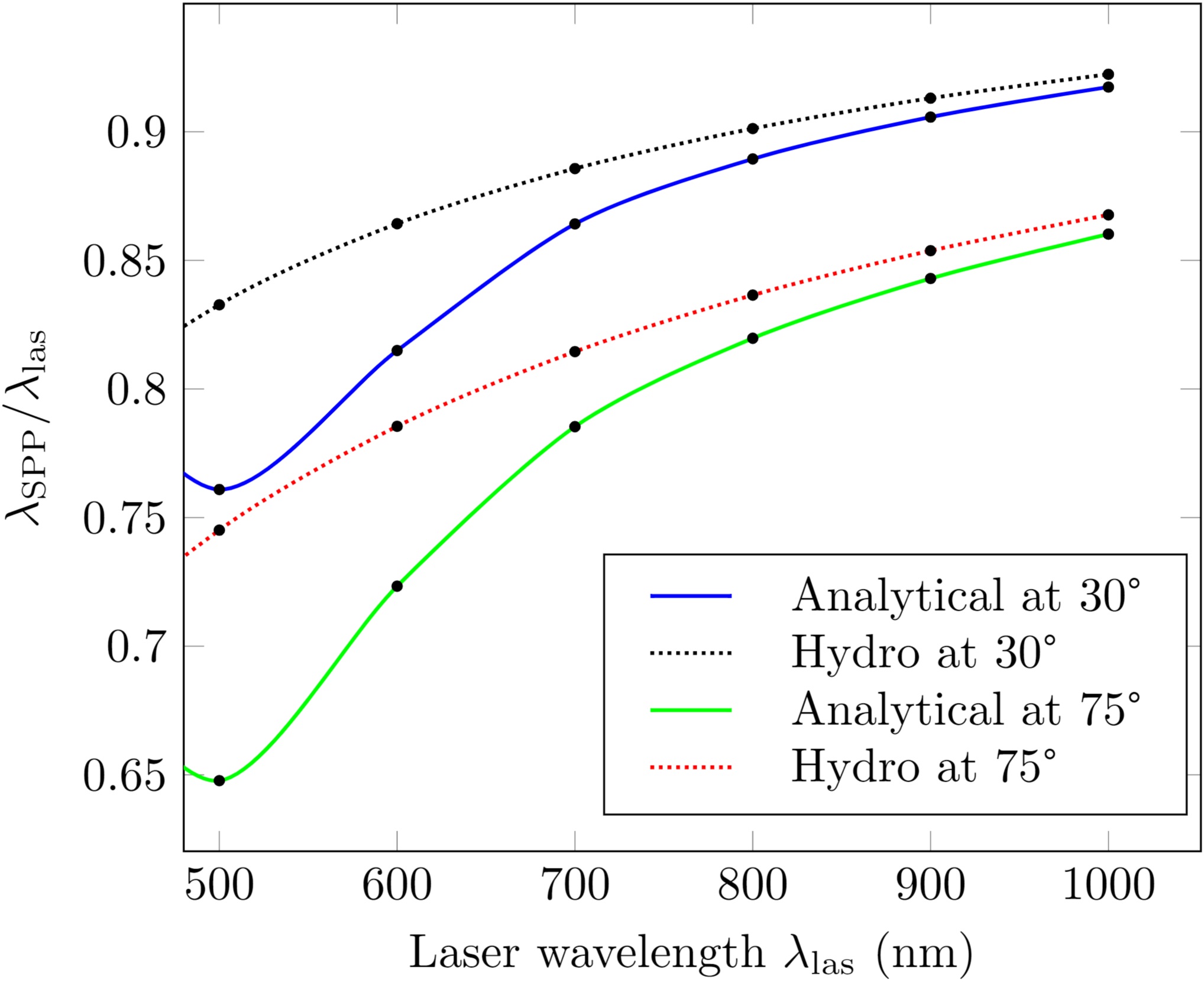}
\caption{Comparison of the ratio $\lambda_{\textrm{SPP}}/ \lambda_{\textrm{las}}$ between predictions from the analytical and hydrodynamic dielectric functions respectively, with respect to the laser wavelength.}
\label{lamdasppfig}
\end{figure}
Fig. \ref{lamdasppfig} shows a comparison between the ratio of $\lambda_{\textrm{SPP}}$ and $\lambda_{\textrm{las}}$, calculated using the dielectric function from Hydrodynamic model and an analytical dielectric function. This analytical dielectric function $\epsilon_{\textrm{Au}}$ has been fitted with experimental data for gold, and takes interband transitions into account using the critical points method (see ref. \cite{etchegoin2006analytic}). We can see from the figure that the predictions from both dielectric functions converge to the same values and have a good agreement at long wavelengths, however for shorter wavelengths, the hydrodynamic model is no longer reliable, and fails to reproduce some properties of the electronic system, such as the dip around 500 nm, which represents photo-excitation processes of d-band electrons of gold. This is due to the fact that the hydrodynamic model, does not yet include interband transitions. However, these effects can be incorporated as polarisation currents using the Lorentz oscillators, see for instance \cite{mcmahon2010calculating}.

\section{Conclusions}\label{conclusion}

 A two temperature (plus) plasmon model was derived from conservation equations of the electronic system. The model consists of three equations for energy conservation of the three subsystems (thermalized electrons, SPP and the lattice), coupled with four other equations governing the dynamics of the electromagnetic field created by the SPP. This field has been shown to have a great influence on the total energy redistribution in the system. The model solves the field equations of these collective oscillations without imposing an arbitrary functions to shape the SPP fields. The resulting interference between SPP and the laser source periodically shapes the lattice temperature profile above the ablation threshold. We have also shown, an overall agreement at long wavelengths, between the predictions of the SPP periodicity between the Hydrodynamic dielectric function and the one fitted from experimental data. Finally, we would like to point out some of the limitations of the TTM+P. This model cannot be applied for electronic energies comparable or higher than the Fermi energy. Also, at short laser wavelengths, the hydrodynamic model's predictions become inaccurate since it needs to integrate interband effects. 
We further mention the absence of Landau damping in the collision term between the SPP and the electrons. As a consequence, the model does not account for processes like electron-hole pair creation and electron photo-emission, thus more accurate terms could prove to be beneficial to improve the TTM+P.
\\ This model, at its current stage, takes the plasmonic system as a separate degree of freedom. This enables us to not only derive the SPP electromagnetic fields, but also investigate the interaction between the different subsystems (i.e. electrons, plasmons and lattice). Also, extending the theory to higher orders of perturbations (second and third orders), one is able to study nonlinear effects occurring in the plasmonic system (see for instance \cite{kravtsov2018enhanced}). Overall, we hope that in the future, our model can be easily adapted for different use cases, and provide good quantitative results for plasmonic investigations.

\section{Declaration of Competing Interest}
The authors declared that there is no conflict of interest.

\section{Acknowledgments}
We wish to acknowledge the financial support of the Deutsche Forschungsgemeinschaft for the project GA465/15-2 as well as for the project RE1141/14-2.

\bibliography{mybibfile}

\end{document}